\documentclass[showpacs,twocolumn,floatfix,amssymb,amsmath,aps,prl]{revtex4}
\usepackage[dvips]{graphicx}
\usepackage[ps2pdf,colorlinks=false,pdfpagemode=UseNone]{hyperref}
\usepackage{latexsym,epsfig,bm,times,psfrag,subfigure} 
\usepackage{dcolumn}
\begin{document}

\renewcommand{\r}{{\bf{r}}}
\renewcommand{\S}{{\mathcal{S}}}
\renewcommand{\k}{{\bf{k}}}
\renewcommand{\arcsin}{\operatorname{arc\,sin}}
\renewcommand{\arccos}{\operatorname{arc\,cos}}
\newcommand{\arctanh}{\operatorname{arc\,tanh}}
\renewcommand{\ap}{{a^{\perp}}}
\newcommand{\gec}{c}
\newcommand{\g}{g}
\newcommand{\phdagg}{^{\phantom{\dagger}}}
\newcommand{\phid}{\phi^{\dagger}}
\newcommand{\phipd}{\phi}
\newcommand{\psid}{\psi^{\dagger}}
\newcommand{\psipd}{\psi^{\phantom{\dagger}}}
\newcommand{\phis}{\phi^{\ast}}
\newcommand{\phips}{\phi^{\phantom{\ast}}}
\newcommand{\psis}{{\psi^{\ast}}}
\newcommand{\psips}{\psi^{\phantom{\ast}}}
\renewcommand{\psips}{\psi}
\renewcommand{\psipd}{\psi}
\renewcommand{\phips}{\phi}
\renewcommand{\phipd}{\phi}
\newcommand{\phio}{\overline{\phi}}
\newcommand{\phipo}{{\phi}}
\newcommand{\psio}{\overline{\psi}}
\newcommand{\psipo}{{\psi}}
\newcommand{\order}[1]{\mathcal{O}(#1)}
\newcommand{\mb}[1]{\mathbf{#1}}
\newcommand{\be}{\begin{equation}}
\newcommand{\ee}{\end{equation}}
\newcommand{\ba}{\begin{align}}
\newcommand{\ea}{\end{align}}
\newcommand{\s}{{\sigma}}
\newcommand{\si}{{\sigma}}
\newcommand{\rp}{{\bf{r}^{\prime}}}
\newcommand{\rf}{{\bf{r}}_f}
\newcommand{\ri}{{\bf{r}}_i}
\newcommand{\q}{{\bf{q}}}
\newcommand{\qp}{{\bf{q^{\prime}}}}
\newcommand{\kp}{{\bf{k^{\prime}}}}
\newcommand{\p}{{\bf{p}}}
\newcommand{\pp}{{\bf{p^{\prime}}}}
\newcommand{\ham}{\mathcal{H}}
\newcommand{\G}{{\mathcal{G}}}
\newcommand{\GF}{{\mathcal{\tilde{G}}}}
\newcommand{\PiOF}{{{{\tilde{\Sigma}}}}}
\newcommand{\PiO}{{{{\Sigma}}}}
\newcommand{\epsl}{\epsilon}
\newcommand{\uu}{\uparrow}
\newcommand{\dd}{\downarrow}
\newcommand{\Tr}{{\rm{Tr}}}
\newcommand{\orot}{\Omega_{r}}
\renewcommand{\orot}{\omega}
\newcommand{\op}{\Omega_{\perp}}
\newcommand{\oz}{\Omega_{z}}
\newcommand{\barg}{\overline{g}}
\newcommand{\sech}{\operatorname{sech}}
\newcommand{\espilon}{\epsilon}
\newcommand{\OT}{\overline{\Omega}}
\newcommand{\OTO}{\tilde{\Omega}_0}
\newcommand{\az}{{a^{z}}}
\newcommand{\xb}{{\overline{x}}}
\newcommand{\yb}{{\overline{y}}}
\newcommand{\zb}{{\overline{z}}}
\newcommand{\zo}{{z_0}}
\newcommand{\zp}{{z_+}}
\newcommand{\zm}{{z_-}}
\newcommand{\Qo}{{Q^{\dagger}_0}}
\newcommand{\Qp}{{Q^{\dagger}_+}}
\newcommand{\Qm}{{Q^{\dagger}_-}}
\newcommand{\Qpo}{{Q^{\phantom{\dagger}}_{0}}}
\newcommand{\Qpp}{{Q^{\phantom{\dagger}}_{+}}}
\newcommand{\Qpm}{{Q^{\phantom{\dagger}}_{-}}}
\newcommand{\epp}{\overline{\epsilon_0}}
\newcommand{\OTN}{\OT}
\newcommand{\hd}{\overline{\delta}}
\newcommand{\hmu}{\overline{\mu}}
\newcommand{\hT}{\overline{T}}
\newcommand{\od}{\overline{\delta}_0}
\newcommand{\omu}{\overline{\mu}_0}
\newcommand{\omc}{\omega_{c2}}
\newcommand{\oT}{\overline{T}_0}
\newcommand{\eF}{{\epsilon}_{F}}
\newcommand{\eFF}{{\epsilon}^{0}_{F}}
\newcommand{\eee}{\epsilon^{(F)}_{i=\{ i_0,I,i^z=0\}}}
\newcommand{\epso}{\overline{\epsilon}_i}
\newcommand{\II}{\delta(I_I)}
\newcommand{\IH}{\delta(I_h)}
\newcommand{\Deltao}{\overline{\Delta}}
\newcommand{\deltap}{\delta}
\newcommand{\taup}{\tau^{\prime}}
\renewcommand{\log}{\ln}
\title{Superfluid transition in a rotating resonantly-interacting
  Fermi gas} 
\author{ Martin Y.\ Veillette, Daniel E.\ Sheehy, Leo Radzihovsky and 
Victor Gurarie}
\affiliation{Department of Physics, University of Colorado, Boulder,
 Colorado 80309}
\date{July 28, 2006} 
\pacs{03.75.Fi, 03.75.Lm, 05.30.Jp}

\begin{abstract}
  
  We study a rotating atomic Fermi gas near a narrow s-wave Feshbach
  resonance in a uniaxial harmonic trap with frequencies
  $\Omega_\perp$, $\Omega_z$. Our primary prediction is the
  upper-critical angular velocity, $\omc (\delta,T)$, as a function of
  temperature $T$ and resonance detuning $\delta$, ranging across the
  BEC-BCS crossover. The rotation-driven suppression of superfluidity
  at $\orot_{c2}$ is quite distinct in the BCS and BEC regimes, with
  the former controlled by Cooper-pair depairing and the latter by the
  dilution of bosonic molecules. At low $T$ and
  $\Omega_z\ll\Omega_\perp$, in the BCS and crossover regimes of
  $0 \lesssim \delta \lesssim  \delta_c$, $\omc$ is implicitly given by $\hbar
  \sqrt{\omc^2 +\Omega_\perp^2}\approx 2\Delta \sqrt{\hbar
    \Omega_\perp/\eF}$, vanishing as
  $\omc\sim\Omega_\perp(1-\delta/\delta_c)^{1/2}$ near
  $\delta_c\approx 2\eF +
  \frac\gamma2\eF\ln(\epsilon_F/\hbar\Omega_\perp)$ (with $\Delta$ the
  BCS gap and $\gamma$ the resonance width), and extending the bulk result
  $\hbar\omc\approx 2\Delta^2/\eF$ to a finite number of atoms in a
  trap. In the BEC regime of $\delta < 0$ we find
  $\omc\to\Omega^-_\perp$, where molecular superfluidity can only be
  destroyed by large quantum fluctuations associated with comparable
  boson and vortex densities.

\end{abstract}

\maketitle


Recent advances in atomic gases near a Feshbach resonance (FR) have
led to the realization of resonantly-paired atomic
superfluids~\cite{Regal2004,Zwierlein2004}.  The proximity to a FR
allows a tunability of the pairing interaction, thereby permitting
unprecedented access to fermionic superfluidity ranging from a
weakly-paired BCS regime to a strongly-paired molecular BEC regime.

A fundamental aspect of a superfluid is its non-classical response to
an imposed rotation. Unable to exhibit rigid body rotation, a
superfluid rotates by nucleating a vortex array with density $n_v =
m\omega/\pi\hbar$ set by the rotation rate $\omega$.  Although
considerable progress has been made in elucidating the properties of
rotating superfluids in the BEC regime (with bosonic
atoms)~\cite{Matthews1999,Abo-shaeer2001,sr}, considerably less is
understood for a resonantly-paired trapped superfluid in the BCS and
crossover regimes. Recent spectacular experiments by Zwierlein, {\it
et al.}~\cite{Zwierlein2005} provide strong motivation for a study of
these regimes.

Additional motivation is provided by the relation of a rotating
superfluid to a type-II superconductor in a magnetic field
~\cite{Abrikosov1957,Gorkov1959b,Helfand1966}, with the Coriolis force
$2 m \dot{ \bf r } \times {\mbox{\boldmath{$\omega$}}}$ in the former
corresponding to the Lorentz force in the latter, with the
identification of $-eB/c$ with $2m\omega$. Although significant
insight can be gained from this connection, it is limited to the BCS
regime and does not include important ingredients that are unique to a
rotating superfluid. These include: the absence of screening, the
centrifugal force, the trap potential, the fixed number of atoms, and
the tunable resonant pairing interaction, all of which are absent in
the analogous superconductor problem. In the latter a concomitant
Zeeman field also appears. It can be effectively introduced into the
atomic problem (but we will not do so here) by imposing a difference
in the number of the two pairing atomic (hyperfine state)
species~\cite{Sheehy2006}.

\begin{figure}[ht]
\begin{center}
\begin{picture}(8.6, 5.5)(0.0, 0.0)
\put(0,
0){\includegraphics[width=8.6cm]{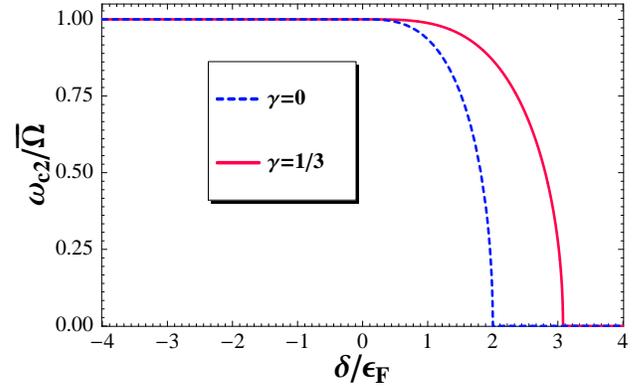}}
\end{picture}
\caption{(Color online) The upper critical rotation rate
  $\orot_{c2}(\delta,T \rightarrow 0)$ as a function of the detuning
  $\delta$ in an isotropic trap with $10^7$ atoms, and for two FR widths $\gamma = 0,1/3$.
 $\omc$ vanishes for $\delta$
  greater than $\delta_c\approx 2\eF
  +\frac\gamma2\eF\ln(\epsilon_F/\hbar\Omega_\perp)$.}
\label{Figure1}
\end{center}
\end{figure}

\begin{figure}[ht]
\begin{center}
\begin{picture}(8.6,5.5 )(0, 0)
\put(4.9,3){\includegraphics[width=3.6cm]{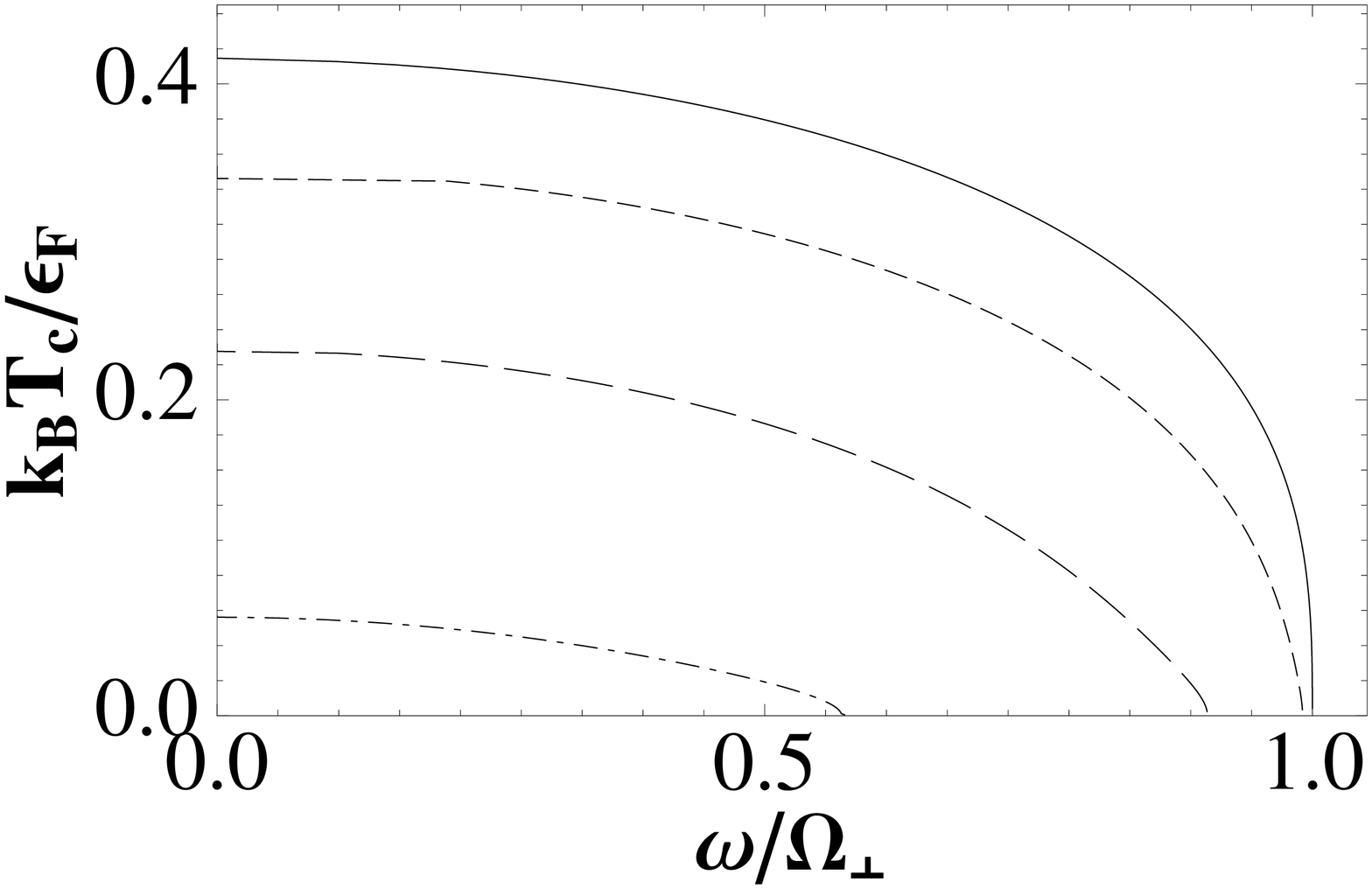}} 
\put(0,0){\includegraphics[width=8.6cm]{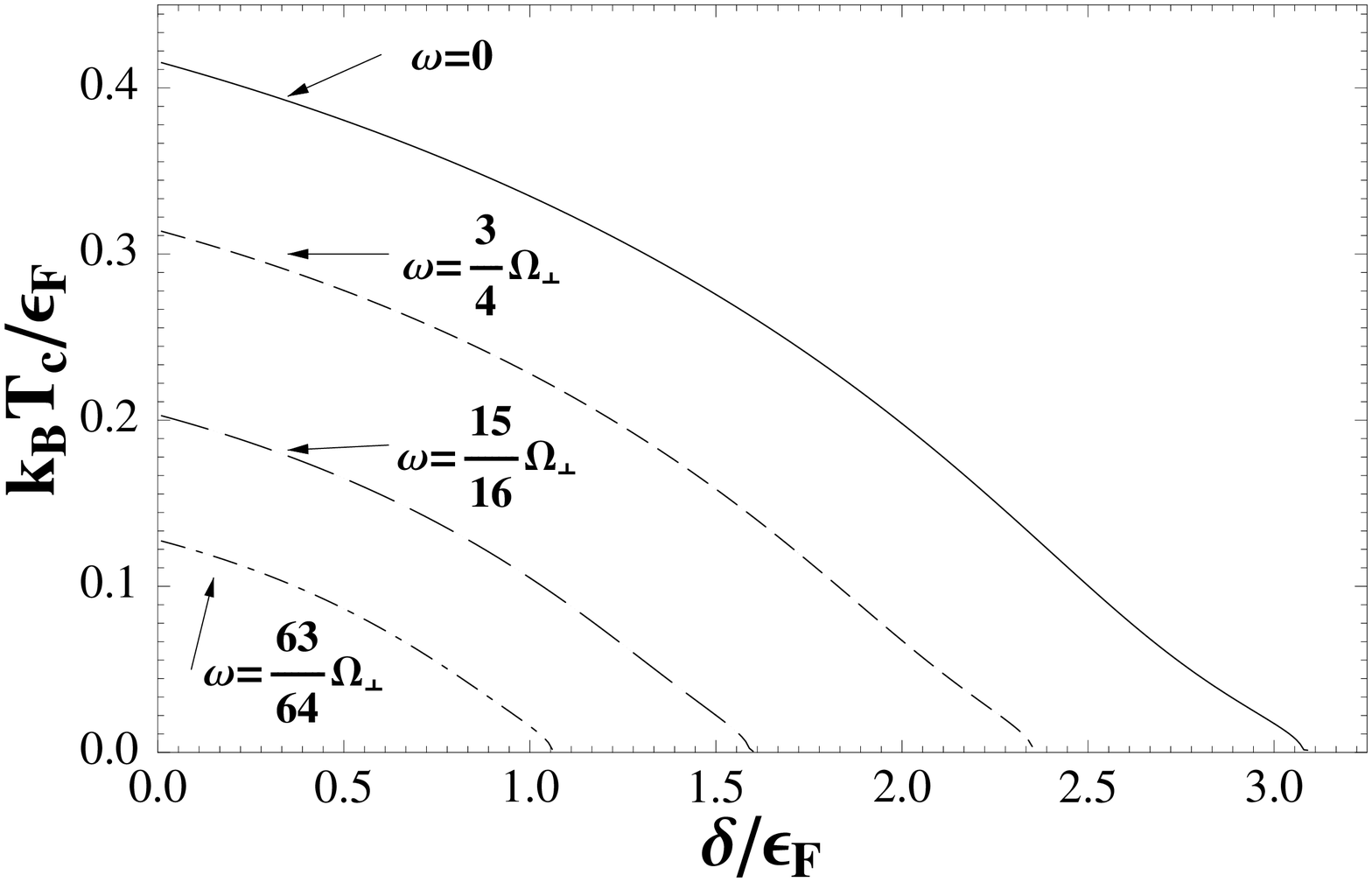}} 
\end{picture}
\caption{Main: $T_c(\delta,\orot)$ in a spherical trap for a range of rotation rates $\orot$ as a function of $\delta$. Inset: $T_c(\delta,\orot)$ for a range of detuning $\delta=\{0,0.9,1.8,2.7\} \eF$  as a function of $\orot$ with the top line being for $\delta=0$. In both cases, $N=10^7$ and $\gamma=1/3$.}
\label{Figure2}
\end{center}
\end{figure}

In this Letter we study the effect of an imposed rotation $\orot$ on a
trapped (with axially symmetric trap frequencies $\op$ and $\oz$ and
$\OT \equiv(\oz\op^2)^{1/3}$ ) resonantly-paired superfluid near a
{\em narrow} FR (i.e., the width of the resonance, $\Gamma$, is
smaller than the Fermi energy $\eF=(3 N)^{1/3}\hbar \OT \equiv\hbar^2
k_F^2/m$, $\gamma \equiv\sqrt{\Gamma/\eF} \ll 1$), tuned through the
BEC-BCS crossover\cite{Andreev2004}. Our primary prediction is the
upper-critical angular velocity $\omc(\delta,T)$ as a function of FR
detuning $\delta$ and temperature $T$, illustrated for $T \rightarrow
0$ in Fig.~\ref{Figure1} and for a range of $\delta$ as a function of
$T$ in Fig.~\ref{Figure2}.

For low $T$ and $\Omega_z\ll\Omega_\perp$, in the BCS and crossover
regimes of $0 \lesssim \delta < \delta_c$ we find 
$\omc$ to be {\em
  implicitly} given by $\hbar\sqrt{\omc^2 +\Omega_\perp^2}\approx 2
\Delta_{\omc} \sqrt{\hbar \Omega_\perp/\eF^{\omc}}$, controlled by the
discreteness of the rotated trap spectrum cutting off the weak Fermi-surface
Cooper pairing characterized by the BCS gap
$\Delta_{\omega=0}\equiv\Delta(\eF)=8e^{-2} \eF
\exp[-{\delta-2\eF\over\sqrt{\Gamma\eF}}]$.  In this regime, for
$\frac\gamma2\ln(\epsilon_F/\hbar\Omega_\perp)\gg 1$, this leads to
\begin{eqnarray}
\label{omegac2result1}
\hspace{-0.7cm}
\omc^{T=0}(\delta)&\approx&
\begin{cases}
  \sqrt{6}\Omega_\perp\sqrt{1-{\delta\over\delta_c}}, &
  \text{for $\delta\rightarrow\delta_c^-$,}\cr
  \Omega_\perp\sqrt{1 -({\delta\over\delta_c})^6}
    & \text{for $0 \lesssim \delta\ll\delta_c$,}\cr
\end{cases}
\end{eqnarray}
where the behavior near $\delta_c\approx 2\eF +
\frac\gamma2\eF\ln(\epsilon_F/\hbar\Omega_\perp)$ corresponds to a
vanishing of $T_c^{\omega=0}$ when the BCS condensation energy
$\Delta^2/\epsilon_F$ drops down to the trap level spacing
$\hbar\Omega_\perp$.  At low detuning $\delta\ll \delta_c$, $\omc$
rises up to but is limited below $\Omega_\perp$ by the implicit
dependence entering through $\eF^{\omega}=\eF
(1-\omega^2/\op^2)^{1/3}$ and $\Delta_\omega=\Delta(\eF^\omega)$ due
to the centrifugal and Coriolis forces reducing the effective trap
potential and atom density.  In this limit, we recover the bulk result
$\hbar\omc\approx 2\Delta_{\omega_{c2}}^2/\eF^{\omega_{c2}}$,
corresponding to Gorkov's (fixed chemical potential) prediction for
type-II superconductors~\cite{Gorkov1959b}.

In the opposite limit of
$\frac\gamma2\ln(\epsilon_F/\hbar\Omega_\perp)\ll 1$, $\delta_c\approx
2\eF$ and $\omc^{T=0}$ reduces to
\begin{equation}
\omc^{T=0}(\delta)\approx\omega_*(\delta)
=\op \sqrt{1-(\delta/2\eF)^3}, \mbox{{ for } } 0 < \delta \lesssim 2\eF,
\label{omegaStar}
\end{equation}
This also smoothly matches onto the $\omc\to\Omega^-_\perp$ behavior in
the BEC regime of $\delta < 0$, where molecular superfluidity can only
be destroyed by large quantum fluctuations associated with comparable
boson and vortex densities, where it undergoes transitions to a variety of
bosonic quantum Hall states~\cite{Wilkin1998}.

As can be seen in Fig.~\ref{Figure1} and Eq.~(\ref{omegac2result1}), a
distinction between BCS ($\delta\gg2\eF$) and crossover ($0<\delta
\lesssim 2\eF$) regimes is not reflected in $\omc(\delta)$, controlled
throughout by weak Cooper pairing. However, $T_c(\omega)$
(Fig.~\ref{Figure2}) does distinguish between these regimes,
exhibiting a point of inflection at a scale $\omega_*(\delta)$ in the
crossover (but not in the BCS) regime. At a rotation rate $\omega <
\omega_*(\delta)$ (nonzero only in the crossover regime) a finite
number of molecular bosons $N_B(\omega)$ is present and $T_c(\omega)$
is set by $T^{\rm BEC}_c[N_B(\omega,\delta)]$. At a higher rate,
$\omega_*(\delta)<\omega <\omc(\delta)$ (as a consequence of an
increased atomic density of states) $\eF^\omega$ drops below
$\delta/2$ and $T_c(\omega)$ is determined by an exponentially reduced
bulk BCS value $T^{\rm BCS}_{\rm
  c}(\eF^\omega)=\frac{e^c}{\pi}\Delta(\eF^\omega)$ ($c\approx
0.577$ is the Euler-Mascheroni constant).

In the BCS regime $(\delta > 2 \eF )$, at slow rotation
$\omega\ll\omc^{T=0}$, we find
\begin{align}
&\frac{T_c(\orot)}{T^{\rm BCS}_{\rm c,bulk}} = 
 1-\frac{7\zeta(3)}{12 \pi^2 } 
\frac{\eF  \hbar  \left( \op^2 +\orot^2 \right)}
{\op \left( k_B T^{\rm BCS}_{\rm c,bulk}\right)^2} + \dots,
\label{Tcomega}
\end{align} 
with the trap responsible for the quadratic in $\orot$ suppression of
$T_c$ in the BCS regime, contrasting to the linear $|\orot|$
suppression in a superconductor~\cite{Abrikosov1957}.  In the BEC
regime, the rotation suppresses $T_c$ according to
\begin{align}
T_c(\orot) \approx T^{{\rm BEC}}_{c} \left(1- \orot^2/\op^2
\right)^{1/3}, \mbox {{ for } } &\deltap
\lesssim 0 .
\label{Tbecomega}
\end{align}
We now sketch the derivation of these results, delaying details to a
future publication~\cite{MSRGunpublished}.

A gas composed of two hyperfine species (labeled by $\sigma=\uu,\dd$)
of fermionic atoms interacting through a tunable (via a \lq\lq
bare\rq\rq\ detuning $\delta_0$) s-wave FR, corresponding to a
closed-channel diatomic molecular state, is well-characterized by a
Hamiltonian $\ham=\ham_F+\ham_B +\ham_g$,
\cite{Timmermans1999,Holland2001,Andreev2004} where
\begin{align}
\ham_F & = \sum_{\s} \int d^3 \r \: 
\psid_\s (\r) 
\left[ -\frac{\hbar^2 \nabla^2 }{ 2 m} + V(\r) - \mu - \orot L^z \right] 
\psipd_\s (\r), \notag \\
\ham_B & = \int d^3 \r \: 
\phid(\r) 
\left[ -\frac{\hbar^2 \nabla^2 }{ 4 m} + 2 V(\r) - 2\mu +\delta_0 - \orot L^z \right] 
\phipd(\r), \notag \\
\ham_g & = g \int d^3 \r \! \left( \phid(\r) \psipd_\uu (\r) \psipd_\dd (\r)
\!+\!\psid_\dd (\r) \psid_\uu (\r) \phipd(\r) \right),
\end{align}
written in the frame rotating with $\orot\hat{z}$. Here $\psid_\s(\r) $ is an
atomic creation operator in the hyperfine state $|\s \rangle$ whereas $\phid
(\r)$ is the molecular creation operator. Above, 
 $m$ is the fermion mass, $\mu$ the chemical potential, 
$g$ the atom-molecule interconversion amplitude, 
$L^z= -i \hbar \hat{z} \cdot \left( \r
 \times \nabla_\r \right)$ the angular momentum operator, and $
V(\r)= \frac12 m \left( \op^2 (x^2+y^2) + \oz^2 z^2 \right)$ is
the atomic trap potential.

To determine $\omc(\delta,T)$ we look for an instability of the
normal state to a superfluid state. To this end we consider system's
partition function $Z=\Tr [e^{-\beta \ham}]$, $(\beta=1/k_B T )$ and integrate out Fermi fields
perturbatively in the FR coupling $g$, valid when the corresponding
dimensionless coupling $\gamma= \frac{1}{4 \pi^2}
\frac{g^2}{\sqrt{\eF}} \left(\frac{2 m}{\hbar^2} \right)^{3/2} \ll 1$.
We obtain 
$Z=\int D \phio D \phipo
\exp(-{S_\phi}/\hbar)$ with (defining $\hat{h}_B= -\hbar^2\nabla^2/4m +
2V(r)-2\mu +\delta_0 -\orot L_z)$,
\begin{align}
{S_\phi} & =- \hbar \ln \left[ Z^{(0)}_{F} \right] + \! \int_0^{\beta
 \hbar} \! d \tau \!
\int \!
d^3 \r \; 
\phio(\r,\tau) \left( \hbar \partial_\tau + \hat{h}_B \right)
\phipo(\r, \tau) \notag \\
&+\frac{1}{\hbar} \! \int_0^{\beta \hbar} \! d \tau d \taup 
\! \int \! d^3 \r d^3 \rp \; 
\phio(\r,\tau) \PiO (\r,\rp,\tau-\taup)
\phipo(\rp, \taup), 
\end{align}
where $\PiO(\r,\rp,\tau)= -g^2 \G^{(0)}_F(\r,\rp,\tau)
\G^{(0)}_F(\r,\rp,\tau)$ is the molecular (Cooper pair) self-energy
arising from molecule fluctuations into a pair of atoms governed by
the free fermion propagator $\G^{(0)}_F(\r,\rp,\tau)= -\langle
\psipo_{\s} (\r,\tau) \psio_{\s} (\rp,0) \rangle_{0} $ and
$Z^{(0)}_{F}= {\rm Tr} \left[e^{-\beta \ham_{F}} \right]$ is the
partition function of free fermions.  Higher-order corrections in $g$
lead to the inclusion of molecular interactions induced by the atomic FR,
small for $\gamma \ll 1$.
 
The instability to the superfluid state at $\omc(\delta,T)$ is
determined by the vanishing of the lowest eigenvalue of the inverse of
the effective molecular propagator $\GF_B^{-1}(\r,\rp;i
\omega_{\ell})|_{i\omega_{\ell}=0} \equiv \GF_B^{(0) -1} (\r, \rp ;i
\omega_{\ell})- \PiOF (\r,\rp, i \omega_{\ell})|_{i \omega_{\ell}=0}$,
where $\G_B^{(0)}(\r,\rp, \tau) = -\langle \phipo (\r,\tau) \phio
(\rp,0) \rangle_{0} $ is the free molecular propagator and
$\omega_{\ell}=2 \ell \pi/(\beta \hbar), (\ell \in \mathbb{Z})$ is the
bosonic Matsubara frequency\cite{Thouless1960}. To lowest order in
$g^2$, $\GF_B^{-1}(\r,\rp;0)$ is diagonalized by eigenstates of
$\GF_B^{(0)-1}(\r,\rp;0)$ (valid for a narrow resonance), that are
simply rotated trap eigenstates $\phi_{\bf n}(\r)$ defined by $
(-\hbar^2\nabla^2_{\r}/4m + 2V(\r)-\orot L_z) \phi_{\bf
  n}(\r)=\epsilon^{0}_{\bf n} \phi_{\bf n}(\r)$, with eigenvalues
(${\bf n}=\{n_0,n,n_z\}$)
\cite{footnote1} 
\be \epsilon^0_{{\bf n}}= \hbar \op
\left(2 n+1 \right) -2 \hbar \orot n_z +\hbar \oz \left( n_0+ 1/2
\right), 
\ee 
where $n_0 \in \mathbb{N}, n \in \frac12 \mathbb{N}$ and
$n_z=\left\{-n,-n+1, \dots, n \right\}$ are the axial, radial and
angular momentum quantum numbers.  The single-particle fermionic and
bosonic spectra are then given by $\epsilon_{\bf
  n}^{0(F)}=\epsilon^0_{\bf n}-\mu$ and $\epsilon_{\bf
  n}^{0(B)}=\epsilon^0_{\bf n}+\delta_0 -2\mu$, respectively.  In the
weakly-interacting limit the condensation is therefore into ${\bf n}=
{\bf 0} \equiv \{0,0,0\}$, a nonrotating bosonic eigenmode localized
at the center of the trap \cite{footnote1}.

Projecting the self-energy operator $\PiOF(\r,\rp;0)$ onto the condensate
$\phi_{\bf 0}(\r)$ gives $\omc$ as the solution of the Thouless criterion
$\epsilon^{(B)}_{\bf 0}=\epsilon^{0(B)}_{\bf 0} - \int d^3 \r d^3 \rp
\phis_{\bf 0}(\r) \PiOF(\r,\rp;0) \phips_{\bf 0}(\rp) $, explicitly given by
\begin{widetext}
  \be 1=\frac{g^2}{\deltap -2\mu} \left\{ \sum_{{\bf n} ,{\bf n'}}
    |R_{{\bf n},{\bf n^{\prime}}}|^2 \frac{ \tanh \left( \beta
        \epsilon^{0 (F)}_{\bf n}/2 \right) + \tanh \left( \beta
        \epsilon^{0 (F)}_{\bf n^{\prime}}/2 \right) } {2 \left(
        \epsilon^{0 (F)}_{\bf n} +\epsilon^{0 (F)}_{\bf n^{\prime}}
      \right)} - \int \!  \frac{d^3 k}{(2 \pi)^3} \frac{m}{\hbar^2
      k^2} \right\},
\label{TE}
\ee 
where we have expressed the \lq\lq bare\rq\rq\ detuning, $\delta_0$,
appearing in the Hamiltonian in terms of the physical detuning
$\deltap=\delta_0- g^2 \int \! \frac{d^3 k}{(2 \pi)^3}
\frac{m}{\hbar^2 k^2}$ arising in the two-body s-wave scattering
measurement through the scattering length $a_s = -\frac{m g^2}{4 \pi
  \hbar^2 \deltap}$. The matrix element $R_{{\bf n},{\bf
    n^{\prime}}}\equiv 2^{-3/4}\int d^3 \r \phi_{\bf 0 }(\r) \phi_{\bf
  n}^*(\r/\sqrt{2}) \phi_{\bf n^{\prime}}^*(\r/\sqrt{2})$ is given by
\be
|R_{{\bf n},{\bf n^{\prime}}}|=\frac12\left(\frac{m \OT }{2 \pi
    \hbar}\right)^{3/4}
\frac{[1 +(-1)^{n_0+n_0^{\prime}}](n_0+n_0^{\prime})!}{2^{(n_0+n_0^{\prime})}
 \left(\frac{n_0+n_0^{\prime}}{2} \right)! \sqrt{n_0!n_0^{\prime}!}} 
\frac{(n+n^{\prime})!}{2^{n+n^{\prime}}}
 \frac{\delta_{n_z+n_z^{\prime},0}}{\sqrt{(n-n_z)!(n+n_z)!(n^{\prime}+n_z^{\prime})!(n^{\prime}-n_z^{\prime})!}}.
\ee
\end{widetext}

Equation~(\ref{TE}) needs to be solved together with the total atom
number equation $N= \beta^{-1} d \ln Z/ d \mu=\int d^3 \r \left( 2
  \langle \phi^{\dagger} \phi \rangle + \langle \psi_\sigma^{\dagger}
  \psi_\sigma \rangle \right)$, fixing $\mu$ in terms of $N$, which to
this order in $g$ is given by $N=N_F^0 + 2 N_B + {\mathcal O}(g^2)$
with $N_F^0 = 2\sum_{\bf n} 1/(e^{\beta \epsilon_{\bf n}^{0 (F)}}+1)$
and $N_B =\sum_{\bf n} 1/(e^{\beta \epsilon_{\bf n}^{(B)}}-1)$.

Analytic analysis of Eq.~(\ref{TE}) is possible due to  considerable
simplifications in the limits of a narrow resonance ($\gamma\ll 1$) and
a Fermi energy that is large compared to the rotating trap level
spacing, $\eF \gg \hbar( \op+\orot)$. The latter ensures that
the fermionic trap level sums are dominated by large quantum numbers ${\bf
  n},{\bf n^{\prime}}$, allowing the Gaussian approximation
$(\frac12)^{n+n^{\prime}}(n+n^{\prime})!/(n!n^{\prime}!)  \approx (\pi
n )^{-1/2} \exp \left[ -(n-n^{\prime})^2/4n \right]$.  Physically this
approximation corresponds to a weak Coriolis force, with the atomic
trajectories nearly straight lines (locally well-characterized by
plane waves) turning by a small fraction of the particle spacing,
$n^{-1/3}$. This reduces the Thouless criterion, Eq.~(\ref{TE}), to \be
\frac{\delta-2 \mu}{\gamma \sqrt{\eF \mu}} = F (\beta \mu)+ G
\left[\beta \mu, \beta \hbar \op \left( 1+
    \frac{\orot^2_{c2}}{\op^2}\right), \beta \hbar \oz \right],
\label{TE1}
\ee
where
\begin{align}
&G(x,a,b) =\int_0^{\infty} \! \int_0^{\infty}
\! dy dp \int_0^1 dz \frac{ e^{-p^2} \sqrt{\frac{y}{\pi x}}
  \tanh(\frac{y-x}{2}) \frac{-1}{y-x}}{
1+\frac{\cosh^2(\frac{y-x}{2})}{\sinh^2 \left(\frac{p}{2}\sqrt{y
        [a+(b-a)z^2]} \right)} } , \notag \\
&G(x\gg1,a,b)= \int_{-\infty}^{\infty} dp
\frac{-e^{-p^2}}{\sqrt{\pi}} \int_0^1 dz \left[c_1
+\psi_0 \left(\frac12+ i Q\right)\right], \\
& F(x) = \int_0^{\infty} dy \sqrt{\frac{y}{4x}} \left[ \frac{\tanh((y-x)/2)}{y-x} -
 \frac{1}{y} \right], \\
&F(x\gg 1)= \ln ( 8 x/\pi )-2
+\gec,
\end{align}
with $Q=p \sqrt{x(a+(b-a)z^2)}/(2\pi)$, $c_1=\ln 4 +\gec$, and
$\psi_0(y)$ the digamma function. Within the same approximation, the
number equation reduces to
\begin{align}
N& \approx 2 \left(\frac{ k_B T}{\hbar \OT[\orot]} \right)^3 \left[ 
-Li_3 \left(-e^{-\beta \epsilon^{0 (F)}_{\bf 0} }\right)+ Li_3 \left(e^{-\beta
 \epsilon_{\bf 0}^{(B)}} \right) \right], 
\label{NUMBER}
\end{align}
where $Li_3(x) = \sum^{\infty}_{k=1} \frac{x^k}{k^3}$ is the
trilogarithm function, with asymptotic forms $Li_3(-e^z)\approx
-z^3/6$ for $z\gg 1$, $Li_3(-e^z)\approx -e^z$ for $z\ll -1$, and
$Li_3(1)\equiv\zeta(3) \approx 1.202 \;$. The first and second terms
are the number of atoms and thermally excited molecules (with the
condensate vanishing at $\omc$), respectively, and
$\OT(\orot)\equiv\OT (1 - \orot^2/\op^2)^{1/3}$ is the effective trap
frequency, reduced by the centrifugal \lq\lq potential\rq\rq.

In the thermodynamic limit $\beta\hbar\Omega_{\perp,z}\ll 1$ and
$\omega=0$, $G(x,0,0)=0$ and (for a narrow FR, $\gamma\ll 1$) we recover the
BCS-BEC crossover with $T_c(\delta)$ ranging from an exponentially
small BCS-regime ($\delta > 2\eF\gg k_B T$) value $k_B T^{{\rm
    BCS}}_{c,\rm bulk} = \frac{8}{\pi} e^{-2 +\gec} \eF \exp
\left[-\frac{\deltap-2 \eF}{\gamma \eF} \right]$ through $T_c^{\rm
  cross.}\approx T_c^{\rm
  BEC}[1-\left({\delta\over2\eF}\right)^3]^{1/3}$ in the crossover
$0<\delta<2\eF$ regime and saturating at $T_c^{\rm
  BEC}=(2\zeta(3))^{-1/3}\hbar\overline{\Omega}N^{1/3}$ for large negative $\delta$.

At finite $\omega$ and low $T$, $\beta\Omega_\perp\gg1$ (specializing
for simplicity to an anisotropic trap
$\beta\Omega_\perp\gg1$, $\beta\Omega_z\ll 1$), using $G(x\gg 1,a\gg 1,
b\ll 1)\approx 1-\frac12 \gec+ \frac12 \ln \left( \frac{\pi^2}{4 x a }
\right) $, we find an {\em implicit} equation $\hbar\sqrt{\omc^2
  +\Omega_\perp^2}=\frac12 e^{1+c/2}\Delta_{\omc} \sqrt{\hbar
  \Omega_\perp/\eF^{\omc}}$ quoted in the introduction. In the BCS and
crossover regimes the implicit $\omc$ dependence enters through $\mu$
as the solution of the number equation, Eq.~(\ref{NUMBER}), that gives
$\mu\approx\eF^{\omega}\approx(3N)^{1/3}\hbar\OT[\orot]$.  The
resulting solution for $\omc$ is illustrated in Fig.~\ref{Figure1},
with asymptotics, controlled by a dimensionless parameter
$\frac\gamma2\ln(\epsilon_F/\hbar\Omega_\perp)$, summarized in
Eqs.(\ref{omegac2result1}, \ref{omegaStar}).

We find that $\omc$ is driven to zero for $\delta > \delta_c\approx
2\eF + \frac\gamma2\eF\ln(\epsilon_F/\hbar\Omega_\perp)$. This
corresponds to a vanishing of $T_c^{\omega=0}$ when the BCS condensation
energy $\Delta^2/\epsilon_F$ becomes comparable to trap level spacing
$\hbar\Omega_\perp$. Equivalently, this condition corresponds to an
oscillator length $a_\perp=\sqrt{\hbar/m\Omega_\perp}$ dropping to the
coherence length $\xi = \hbar v_F/\pi \Delta$.  

In the opposite limit of low detuning,
$\omc(\delta\ll\delta_c)\rightarrow\op^-$, where the centrifugal and
trapping potentials nearly cancel, the system becomes translationally
invariant and we obtain the bulk result $\hbar\omc\approx
8^{-1}e^{2+c}\Delta_{\omega_{c2}}^2/\eF^{\omega_{c2}}$, corresponding
to Gorkov's (fixed chemical potential) prediction for type-II
superconductors~\cite{Gorkov1959b}. This matches onto the BEC regime
$\omc(\delta < 0)\approx\Omega_\perp$, where a purely molecular
superfluid can only be destroyed by a sufficient dilution down to
comparable boson and vortex densities, driving a quantum transition into
a variety of bosonic quantum Hall states~\cite{Wilkin1998}.

At high $T$ (near $T_c^0$) and slow rotation ($\omega\ll\omc^{\rm
  T=0}$) in the BCS regime we expand $G(x\gg 1,a,0)$ in the limit $a x
\equiv \beta^2 \mu \hbar \op (1+\orot^2/\op^2) \ll 1$ (remaining in
the degenerate limit, $\beta\mu=x\gg 1$), obtaining a quadratic
suppression of $T_c$ with rotation and trap frequencies given in
Eq.~(\ref{Tcomega}). In the homogeneous limit $\orot\rightarrow\op^-$,
this reduces to $T_c(\orot)/ T^{{\rm BCS}}_{c,\rm bulk}\approx
1-\frac{7 \zeta(3)}{6 \pi^2} \hbar |\orot| \mu /(k_B T^{{\rm
    BCS}}_{{\rm Bulk}})^2$, giving a linear reduction with $|\orot|$,
as expected from Abrikosov's theory of
$H_{c2}(T)$~\cite{Abrikosov1957}.

In the crossover regime at slow rotation,
$\mu\approx\delta/2<\eF^\omega$ and a finite fraction of the atomic
Fermi sea is bound into a molecular superfluid, giving
$N_F(\orot,\delta)\approx (\delta/2\eF^\omega)^3N=(1/3)(\delta/2\hbar
\OT)^3 /[1-(\orot/\op)^2] < N$. Because the atomic density of states
increases with $\omega$ according to $1/\OT[\orot]^3$ (heading toward
the extensive degeneracy of Landau levels in the $\orot\rightarrow\op$
limit), at sufficiently large $\omega$, $\eF^\omega$ drops below
$\delta/2$. The crossover frequency $\omega_*(\delta)$, defined by
$\eF^{\omega_*}=\delta/2$, corresponds (for a narrow resonance
$\gamma\ll 1$) to a vanishing molecular condensate,
$N_F(\omega_*,\delta)=N$ and is given by Eq.~(\ref{omegaStar}).
$\omega_*(\delta)$ is a lower bound for $\omc(\delta)$ and marks the crossover
between a strongly-paired molecular superfluid, with $T_c\approx
T_c^{\rm BEC}[N_B(\omega,\delta)]$
\begin{equation}
T_c(\omega,\delta)\approx
T_c^{\rm BEC}\left[1-\left({\delta\over2\eF}\right)^3
-{\omega^2\over\Omega_\perp^2}\right]^{1/3}, 
\mbox{{ for } } \omega < \omega_*(\delta),
\label{Tc<omegaStar}
\end{equation}
and a weakly-paired BCS superfluid with $T_c\approx T^{\rm BCS}_{\rm
  c}(\eF^\omega)$ for $\omega_*(\delta)<\omega < \omc(\delta)$. For
$\delta < 0$ all atoms bind into molecules, $N_F(\mu<0)$ is
exponentially small, and the above result crosses over to the BEC value,
$T_c(\omega,0)$, Eq.~(\ref{Tbecomega}), quoted in the introduction.  The
crossover at $\omega\approx\omega_*(\delta)$ (or equivalently at
$\delta\approx 2\eF^{\omega_*}$) can be seen in $T_c(\delta,\omega)$
displayed for a full range of detunings and rotation rates in
Fig.~\ref{Figure2}.

To summarize, we have studied a resonantly-interacting atomic Fermi
gas across a FR and computed $\omc (\delta,T)$ below which the gas
becomes unstable to a rotating paired superfluid.

We thank N. Cooper and V. Galitski for discussions.  We acknowledge
support from NSF DMR-04499521 (VG), DMR-0321848 (MYV, DES, LR).




\begin{thebibliography}{99}
\bibitem{footnote1}
Naively, this is expected to be accurate only for 
$\gamma\eF/\hbar\op \ll 1$. However, in the opposite (thermodynamic)
limit of a shallow trap, the eigenfunctions of $\tilde{\Sigma}(\r,
\r^{\prime};\omega_\ell)$ and $\hat{h}_B$
(Landau level orbitals) become identical, guaranteeing that such
projection in fact gives the exact eigenvalue of  
$\tilde{\G}^{-1}_B$. Consequently, we expect this approximation to be
accurate as long as a far less restrictive condition,
$\gamma \ll 1$, is satisfied.
\end{thebibliography}
\end{document}